\documentclass[prd, twocolumn, superscriptaddress, preprintnumbers,nofootinbib]{revtex4}

\usepackage{amsfonts,amssymb,amsmath}
\usepackage{bm}
\usepackage{bbm}
\usepackage{epsfig}
\usepackage{color}
\usepackage{slashed}

\newcommand{\beq}{\begin{eqnarray}}
\newcommand{\eeq}{\end{eqnarray}}
\newcommand{\nn}{\nonumber}

\newcommand{\bitem}{\begin{itemize}}
\newcommand{\eitem}{\end{itemize}}

\newcommand{\eql}[1]{\label{eq:#1}}
\newcommand{\eq}[1]{(\ref{eq:#1})}
\newcommand{\Eq}[1]{Eq.~(\ref{eq:#1})}
\newcommand{\fig}[1]{\ref{fig:#1}}
\newcommand{\Fig}[1]{Fig.~\ref{fig:#1}}
\newcommand{\Figl}[1]{\label{fig:#1}}
\newcommand{\Tab}[1]{Table~\ref{tab:#1}}
\newcommand{\Tabl}[1]{\label{tab:#1}}
\newcommand{\Ref}[1]{Ref.~\cite{#1}}
\newcommand{\Secl}[1]{\label{sec:#1}}
\newcommand{\Sec}[1]{Sec.~\ref{sec:#1}}
\newcommand{\Appl}[1]{\label{app:#1}}
\newcommand{\App}[1]{Appendix~\ref{app:#1}}

\newcommand{\fr}[2]{{\frac{#1}{#2}}}

\newcommand{\Dee}{\mathrm{D}}
\newcommand{\DD}{\!\mathop{\vbox{\lineskip=0ex\baselineskip=0ex
                   \hbox{\hspace{-.2ex}{\small $\leftrightarrow$}}
                   \hbox{$\hspace{.0ex}\mathrm{D}$}}}\hspace{-.6ex}}

\newcommand{\e}{\mathrm{e}}
\newcommand{\I}{\mathrm{i}}

\newcommand{\lt}{\left}
\newcommand{\rt}{\right}
\newcommand{\avg}[1]{\langle#1\rangle}

\newcommand{\wba}[1]{{\overline{#1}}}
\newcommand{\mbf}[1]{{\mathbf{#1}}}
\newcommand{\mrm}[1]{{\mathrm{#1}}}
\newcommand{\phtm}[1]{{\phantom #1}}

\newcommand{\lsim}{\lesssim}

\newcommand{\al}{\alpha}
\newcommand{\ga}{\gamma}
\newcommand{\Ga}{\Gamma}
\newcommand{\De}{\Delta}
\newcommand{\ep}{\epsilon}
\newcommand{\la}{\lambda}
\newcommand{\La}{\Lambda}
\newcommand{\sg}{\sigma}
\newcommand{\Sg}{\Sigma}
\newcommand{\tha}{\theta}

\newcommand{\cL}{\mathcal{L}}
\newcommand{\cO}{\mathcal{O}}

\newcommand{\C}{\mathrm{c}}
\newcommand{\T}{\mathrm{T}}

\newcommand{\Z}{\mathbb{Z}}
\newcommand{\U}{\mathrm{U}}
\newcommand{\SU}{\mathrm{SU}}

\newcommand{\SUC}{\mathrm{SU(3)_C}}
\newcommand{\SUL}{\mathrm{SU(2)_L}}

\newcommand{\MeV}{~\mathrm{MeV}}
\newcommand{\GeV}{~\mathrm{GeV}}
\newcommand{\TeV}{~\mathrm{TeV}}
\newcommand{\fb}{~\mathrm{fb}}
\newcommand{\s}{~\mathrm{s}}
\newcommand{\ns}{~\mathrm{ns}}
\newcommand{\mm}{~\mathrm{mm}}

\newcommand{\cc}{\text{c.c.}}

\newcommand{\column}[1]{\left(\!\begin{array}{c}#1\end{array}\!\right)}

\begin{document}

\title{LHC Implications of the WIMP Miracle and Grand Unification}

\author{Can K\i l\i\c{c}}
\affiliation{Department of Physics and Astronomy, Rutgers University,
Piscataway, NJ 08854, USA}
\author{Karoline K\"opp}
\affiliation{Department of Physics, Florida State University,
Tallahassee, FL 32306, USA}
\author{Takemichi Okui}
\affiliation{Department of Physics, Florida State University,
Tallahassee, FL 32306, USA}

\begin{abstract}
With the assumptions that dark matter consists of an electroweak
triplet and that the gauge couplings unify at a high scale, we
identify robust phenomenological trends of possible matter contents
at the TeV scale. In particular, we expect new colored states within
the LHC reach that can have Yukawa couplings $\lambda$ to quarks and
the Higgs. We investigate the collider signatures that are characteristic
of all such models by adopting the model with the simplest matter content
as a benchmark. The $\lambda$ couplings are constrained by flavor/$CP$
physics.  In the largest portion of the allowed parameter space the new
colored particles are stable on collider time scales, hence appearing as
R-hadrons, for which there is discovery potential at the \emph{early} LHC
($\sqrt{s}=7\TeV$, $1\fb^{-1}$). Flavor/$CP$ constraints nevertheless do
allow a sizable range of $\lambda$ where the new colored particles decay
promptly, providing a new Higgs production channel with a cross-section
governed by the \emph{strong} interaction. Studying the case of $h \to WW$,
we show that it is possible for the Higgs production from this new channel
to be discovered before that from the Standard Model at the LHC.
\end{abstract}

\preprint{RUNHETC-2010-20}

\maketitle

\section{Introduction}
\Secl{intro}

The existence of dark matter (DM) is arguably the most compelling
evidence for new physics beyond the standard model (SM). Even though
existing data provides little insight into the identity and nature of
the DM particle, a simple and robust candidate is provided by a weakly
interacting massive particle (WIMP), as its relic abundance will be
automatically of the right size when its mass is at the TeV scale (``the
WIMP miracle''). Since the null results from direct DM search experiments
have excluded a WIMP with nonzero hypercharge~\cite{no-Y-for-WIMP}, the
simplest WIMP DM candidate is an $\SUL$-triplet with $Y=0$, denoted as
a $V$ hereafter. The $V$ can be made stable by imposing e.g.~the $\Z_2$
parity $V \to -V$. A careful calculation of the relic abundance of a $V$
is performed in \Ref{minimalDM2}, including non-perturbative effects and
possible co-annihilations, finding that the $V$ mass should be $2.5\TeV$
if the $V$ is spin-0, or $2.7\TeV$ if spin-$1/2$, assuming that the $V$
accounts for the entire missing mass of the universe.

Unfortunately, since the $V$ is heavy and color-neutral, it is virtually
impossible to be directly produced from $pp$ collisions at the Large
Hadron Collider (LHC).
However, $V$ may well be part of a bigger, well-motivated extension of
the SM containing other new particles that can give rise to observable
LHC signals.

In this paper, we adopt gauge coupling unification~\cite{GUT} as a guiding
principle besides WIMP DM.   Although this is still not constraining enough
to point to one single model, it is possible to identify generic, robust
phenomenological trends in such extensions of the SM.  Concretely, we adopt
the $V$ as a dark matter candidate and demand perturbative gauge coupling
unification.%
\footnote{Note that a $V$ is not only the simplest WIMP DM candidate of
all but also the only $\SUL$ multiplet with zero hypercharge that appears
within simple $\SU(5)$ multiplets, $\mbf{5}$, $\mbf{10}$, $\mbf{15}$, and
$\mbf{24}$.}
We assume no extra mass scales other than the unification scale and the TeV
scale dictated by the WIMP miracle. Finally, we assume that new particles,
including the $V$, are all fermions to avoid extra fine-tuning problems
associated with scalar masses. These assumptions readily imply the existence
of additional new particles at the TeV scale, since the gauge couplings in the
SM+$V$ theory do not unify. In particular, we find that there must exist new
colored particles at the TeV scale. We also find that the new colored particles
generically allow Yukawa couplings to quarks and the Higgs.

Split supersymmetry~\cite{splitsusy1, splitsusy2} is a well-studied scenario
also based on WIMP DM and gauge coupling unification, as well as the absence
of supersymmetry at the TeV scale as in our scenario.  There are two major
differences.  First, contrary to one of our assumptions above, split
supersymmetry has an extra threshold between the TeV and unification
scales, where we have all the squarks and sleptons, and most importantly,
the second Higgs doublet, which affects unification. Second, in split supersymmetry,
WIMP DM has to be a nontrivial composition of the higgsinos, wino and/or bino; if we
assume that dark matter in split supersymmetry is a pure wino (i.e.~the $V$) as in
our scenario, gauge coupling unification would not work well (using the same criteria
for precision as we use in \Sec{unification}) unless we have a hierarchy larger than
two orders-of-magnitude between the higgsino and gluino masses.

The existence of new colored particles with Yukawa couplings $\la$ to quarks
and the Higgs suggests the following two scenarios for the LHC. If $\la$ is
sufficiently small, the new colored particles will be collider stable, appearing
as massive stable hadrons (``R-hadrons''). Since R-hadron signals can be quite
spectacular, this is an exciting possibility already for the \emph{early} LHC
run at $\sqrt{s}=7\TeV$ with an integrated luminosity $\approx 1\fb^{-1}$. On
the other hand, if $\la$ is not so small, the new colored particles will decay
promptly via $\la$, with an $\cO(1)$ fraction of their decays containing Higgs
bosons. This is an interesting new production channel for the Higgs boson, where
the size of the cross-section is governed by the \emph{strong} interaction,
potentially making the LHC a ``Higgs factory''.

To perform quantitative benchmark studies of these characteristic
phenomenological features of WIMP DM and unification, we choose as a
simple benchmark model consisting of a DM candidate $V$ and new colored
particles $X$ (to be specified more explicitly later). Among all models
with WIMP DM and unification, this benchmark model contains the smallest
number of new multiplets beyond the SM, but it already exhibits the two
classes of generic collider signatures mentioned above.

Our analysis on this benchmark model will show that, for the range
$m_X = 360$-$650\GeV$, the early LHC phase ($7\TeV$, $1\fb^{-1}$)
should have sufficient discovery potential for the R-hadron case.
For the Higgs factory case, we will lay out an experimental strategy
for the full LHC at $\sqrt{s}= 14~$TeV. This consists of two parts,
the discovery of the $X$ and measurement of $m_X$, and the discovery
of the Higgs bosons from the $X$ decays. We will show that with $10\fb^{-1}$
of data at the LHC (14~TeV), it should be possible to discover the $X$ and
the Higgs bosons from the $X$ decays in the range $300<m_X \lsim 550\GeV$ for
a moderately heavy Higgs (i.e.~decaying to weak gauge bosons).

This paper is organized as follows. In \Sec{unification} we survey
possible extensions of the SM that feature gauge coupling unification
and WIMP DM. In \Sec{benchmark} we describe our simple benchmark model
that phenomenologically represents all such extensions. The couplings
$\la$ of the new colored particles to quarks and the Higgs are expected
to have an upper bound from flavor/$CP$/electroweak constraints, which
is analyzed in \Sec{constraints} using the benchmark model. The collider
signatures of the R-Hadron case and the Higgs factory case are studied in
detail in \Sec{Rhadron} and \Sec{higgsfac}, respectively. In \Sec{conc} we
summarize our analyses and indicate some possible future directions.
In \App{proton} we discuss how proton decay can be avoided in the class
of models we consider in this paper. In \App{higher-dim-op} we comment
that the addition of higher-dimensional operators to our Lagrangian does
not have any impact on our analysis.

\section{Extensions of the SM featuring WIMP DM and coupling unification}
\Secl{unification}
In this section we enumerate possible extensions of the SM that contain
the WIMP DM candidate $V$ and are consistent with gauge coupling unification,
and identify generic features shared by such extensions.  This analysis will
serve as a basis for our choice of a benchmark model in \Sec{benchmark}.

Let us assume unification of the SM gauge group into a simple group (such
as $\SU(5)$) to fix the normalization of hypercharge. The DM candidate $V$
can be embedded into a $\mbf{24}$ of $\SU(5)$, consistent with this assumption.
Moreover, let us assume that all new particles, including the $V$, are spin-$1/2$
fermions in order to avoid unnecessary extra fine-tuning problems associated with
scalar masses besides the notorious existing problem with the Higgs mass. As
calculated in \Ref{minimalDM2}, the fermionic $V$ mass is fixed by the relic
abundance to be $m_V=2.7\TeV$, which we will assume to be the case hereafter.

This cannot be the end of the story, however, because the SM augmented by
only the $V$ is not consistent with gauge coupling unification.  There must
be additional new particles. In principle, these additional particles could
appear anywhere below the unification scale.  However, since we must presume
some underlying dynamics that generates the TeV scale in order for the WIMP
miracle not to be a mere coincidence, we adopt the simplest assumption that the
same dynamics also provides TeV-scale masses to the additional new particles,
with no extra mass scale other than the TeV scale and the unification scale.
We take the new fermions to be vectorlike, because chiral fermions would
require electroweak symmetry breaking to acquire TeV-scale masses, which would
generically lead to dangerously large corrections to precision electroweak
observables, in particular the $\rho$ parameter \cite{4th-gen}.
Let us further restrict ourselves to the case where the vectorlike fermions
can be embedded into the simplest $\SU(5)$ multiplets, $\mbf{5} \oplus \mbf{\bar{5}}$,
$\mbf{10} \oplus \mbf{\wba{10}}$, $\mbf{15} \oplus \mbf{\wba{15}}$ and $\mbf{24}$.  Thus,
we consider%
\begin{center}
\begin{tabular}{lll}
  $Q \sim (\mbf{3},\mbf{2})_{1/6}\,,$  &
  $U \sim (\mbf{3},\mbf{1})_{2/3}\,,$  &
  $D \sim (\mbf{3},\mbf{1})_{-1/3}\,,$ \\
  $L \sim (\mbf{1},\mbf{2})_{-1/2}\,,$ &
  $E \sim (\mbf{1},\mbf{1})_{-1}\,,$   &
  $X \sim (\mbf{3},\mbf{2})_{-5/6}\,,$ \\
  $S \sim (\mbf{6},\mbf{1})_{-2/3}\,,$ &
  $T \sim (\mbf{1},\mbf{3})_{1}\,,$ &
  $V \sim (\mbf{1},\mbf{3})_{0}\,,$ \\
  $G \sim (\mbf{8},\mbf{1})_{0}\,,$
\end{tabular}
\end{center}
as well as the conjugates $Q^\C$, $U^\C$, $\cdots$, $T^\C$, except for $V$ and
$G$, which are real. (Our convention is such that $H$ has the quantum numbers
$(\mbf{1},\mbf{2})_{1/2}$. The SM fermions are denoted by lower-case letters,
$q$, $u^\C$, $d^\C$, $\ell$, $e^\C$.)  As we will see below, this already
provides a sufficient number of candidate models for us to observe generic
trends in extensions of the SM with WIMP DM and gauge coupling unification.

In searching for possible field contents that can lead to unification, there
are various uncertainties that must be taken into account when we ``predict''
the $\SUC$ coupling $\al_3$ in terms of $\al_{1,2}$, which we regard as precise.
First, in our RG analysis, which we perform at the 1-loop level, there is
a threshold ambiguity at $m_V=2.7\TeV$. We estimate this uncertainty by
varying the $\wba{\text{MS}}$ subtraction scale $\mu$ from $m_V/\sqrt{2}$
to $\sqrt{2}\, m_V$. Second, unlike the $V$ mass, the masses of the additional
fermions cannot be fixed a priori and can be anywhere at the TeV scale, from a
few hundred GeV to several TeV. Therefore, we scan over the additional fermions'
masses in the range between $300\GeV$ and $10\TeV$, where for simplicity we
assume a single common mass for all of them.%
\footnote{New charged/colored particles below $300\GeV$ are likely to be already
excluded, as will be illustrated by the analysis of the benchmark model below.}
These two sources of uncertainty each shift our $\al_3$ prediction at the level of a
few times $\De\al_3^{\text{(exp)}}$, the experimental uncertainty in the measurement
$\al_3^{\text{(exp)}} = 0.1184 \pm 0.0007$~\cite{PDG}. We demand that the ``band''
in our $\al_3$ prediction combining these two uncertainties have an overlap with
the band corresponding to $3\De\al_3^{\text{(exp)}}$ (i.e.~$3\sg$). There are also
threshold effects from unspecified GUT physics, but we assume that they are similar
in size to the uncertainties mentioned above and simply neglect them. Finally, we
demand that the coupling at the unification scale be perturbative,
$\al_\text{GUT} < 1$.

A similar analysis was performed in \Ref{splitsusy2}, where the main difference
lies in the treatment of proton decay.  While \Ref{splitsusy2} demands the
unification scale to be higher than $\sim 10^{16}\GeV$ in order to sufficiently
suppress proton decay, we choose to impose a symmetry to forbid proton decay,
and only demand the GUT scale to be higher than $10^5$~TeV (and lower than the Planck scale
$\sim 10^{18}\GeV$) to avoid having to address possible conflicts between GUT physics and
flavor/$CP$ bounds. The use of a symmetry to forbid proton decay requires some model building
at the unification scale, but it has no observational consequences for the TeV-scale physics,
so we leave the model building to \App{proton}. Another difference between our analysis
and that of \Ref{splitsusy2} is that in calculating the running of the gauge couplings,
\Ref{splitsusy2} assumes that all new particles have masses near $m_Z$, while our masses
span a wide range around a TeV, as we described above.

\begin{table}
\begin{tabular}{ccc}
\hline\hline
$2X$         & $2X + D + L$  & $2U + T + E$ \\
$U + S + 2T$ & $2X + 2U + T$ & $3X + L + E$ \\
$3X + G + V$ & $3U + T + L$  & $\cdots$     \\
\hline
\end{tabular}
\caption{\Tabl{unification} Possible combinations of new fermions that, together
with the DM $V$, could lead to unification.  It is understood that the new fermions
are vectorlike, so in writing $X$, $U$, $\cdots$, the presence of their charge
conjugates $X^\C$, $U^\C$, $\cdots$ is also implied, except for the Majorana
fermions $V$ and $G$.}
\end{table}

There are 22 models satisfying the above criteria with no more than 3 types
of multiplets in addition to the $V$ and no more than 3 generations per type.
Particularly simple ones are listed in \Tab{unification}. All the 22 models
share the following features:
\begin{itemize}
\item[(1)]{There exist new colored particles.}
\item[(2)]{The quantum numbers of these new colored particles allow Yukawa
couplings to quarks and the Higgs.}
\end{itemize}
Property (1) is clearly favorable for hadron colliders.  Even better, all the
22 models actually survive even if we restrict the additional fermions' masses
to the range $300\GeV$-$1\TeV$, so they all can be potentially within the LHC
reach.  Property (2) is not satisfied by $S$ and $G$, but they only appear twice
each among the 22 models, and even those models contain other colored particles
that do satisfy property (2).  We therefore identify these properties as robust
LHC implications of WIMP DM and unification.%
\footnote{Recall the crucial role of the WIMP miracle in selecting the TeV scale
as the mass scale for the new particles.}

To assess the robustness of the above features further, one can repeat the exercise
with more conservative estimates on the uncertainties in the prediction of $\al_3$.
For example, if we vary the matching scale $\mu$ from $m_V/2$ to $2m_V$ (with
everything else treated as above), we obtain 63 models, of which there is only
one model ($V + E$) without colored particles,%
\footnote{Actually, a closer inspection reveals that unification favors
the $E$ in the $V+E$ model to be lighter than $\approx 300\GeV$.  The quantum
numbers of the $E$ allow it to decay to $\ell +Z$ in particular, so this model
is already excluded.
}
and only 6 colored models without property (2).
Again, most models survive even if we restrict the search in the
``LHC-accessible'' range $300\GeV$-$1\TeV$; 44 models in total, only one colorless
model, and only one colorful model without property (2).

In the next section, we choose a benchmark model that represents characteristic
phenomenologies of all these models which follow from properties (1) and (2).
We will then use the benchmark model for further, more quantitative analyses
in later sections.

\section{The benchmark model}
\Secl{benchmark}

Given the insights from \Sec{unification}, we select the model with the dark matter
$V$ and two generations of $X\oplus X^\C$ as our benchmark. This is the simplest of
all models with a $V$ featuring unification, having the smallest number of new multiplets
beyond the SM.  But most importantly, the collider phenomenology of this model is
representative of all models identified in \Sec{unification} as far as the LHC
phenomenology is concerned.

The most general renormalizable Lagrangian for the $2X+V$ model consistent with the
$\Z_2$ symmetry $V \to -V$ reads%
\beq
  \cL
    &=& \cL_\text{SM}
       +\wba{V} \bar{\sg}^\mu \I\Dee_\mu V -\fr{m_V}{2} VV  \nn\\
    && \hspace{5ex}
       +\sum_a \Bigl[ \wba{X}_a \bar{\sg}^\mu \I\Dee_\mu X_a
                     +X^\C_a \sg^\mu \I\Dee_\mu \wba{X}{}^\C_a  \eql{benchmark}\\
    && \hspace{5ex}
                     -\Bigl( m_{X_a} X^\C_a X_a^\phtm{\C}
                            +\sum_i\la_{ia} \, H d^\C_i X_a^\phtm{\C} + \cc
                      \Bigr)
               \Bigr]  \,,\nn
\eeq
where $a=1,2$ and $i=1,2,3$ denote generations of the $X$ and $d$-type quark, respectively.
We also use $X_{-1/3}$ and $X_{-4/3}$ to refer to the upper and lower $\SUL$ components of
$X$, respectively, where the subscripts denote the electric charges.

It is technically natural for $\la$ to take any value, but there are obvious phenomenological
constraints.  First, $\la$ has to be nonzero because the $X$s must eventually decay to avoid
cosmological problems. (However, the $X$s could decay via higher dimensional operators.
See \App{higher-dim-op}). Second, $\la$ must be much less than $\cO(1)$ because $\la$ breaks
the $\U(3)^5$ flavor symmetry of the SM, providing new sources of flavor/$CP$ violations
in addition to the SM Yukawa couplings. We will return to flavor/$CP$ constraints in
\Sec{constraints}.

The leading decays of $X$ induced by $\la$ can be most easily understood by the Goldstone
equivalence theorem.  In the limit of keeping only the $X$ mass, the equivalence theorem
tells us that $X_{-1/3}$ will decay as
\beq
  X_{-1/3} \to Z + d  \>,\quad
  X_{-1/3} \to h + d
\eeq
with equal probabilities, where $d$ can be any down-type quark.%
\footnote{The small violation of the equivalence theorem induces additional decays such
as $X_{-1/3} \to W^- + u$, which can be thought of as arising from mixing of the $X$s with
down-type quarks.  However, as we will see in \Sec{constraints}, flavor/$CP$ bounds constrain
such mixings to be tiny, rendering these decay modes negligible.}
Note that the equivalence theorem holds only in this limit. When the finite masses of the Higgs and Z bosons are taken into account, the branching fraction for the Z channel is expected to be somewhat larger due to phase space. For low X masses and/or a heavy Higgs, this can have a significant impact on the phenomenology. Similarly, $X_{-4/3}$ can decay as
\beq
  X_{-4/3} \to W^- + d  \,,
\eeq
which will be the dominant decay mode as long as the corresponding rate can be regarded
as prompt on the collider time scale.  When this rate drops below the displaced-vertex range,
the dominant decay of the $X_{-4/3}$ will be through the weak interaction to an $X_{-1/3}$,
which becomes slightly lighter than $X_{-4/3}$ after electroweak symmetry breaking, as we
will elaborate more in \Sec{Rhadron}.

Therefore, depending on the size of $\la$, we have either of the following collider signatures:
\bitem
\item[(A)]{If $\la$ is sufficiently tiny, the $X$ will be stable on collider time scales,
and upon production it will hadronize into stable massive hadrons (``R-hadrons''). R-hadrons
are easy to observe when they are charged, so $X$ may be discoverable already in the
early LHC run (i.e.~$7\TeV$, $1\fb^{-1}$).
}
\item[(B)]{If $\la$ is not so small (but small enough to satisfy flavor/$CP$ constraints),
the $X$ will decay within the detector, and as we have seen above, roughly a quarter of
$X$s (a half of $X_{-1/3}\,$s) will decay to a Higgs boson (plus a jet). This is an exciting
possibility --- a ``Higgs factory'' --- where the Higgs bosons are produced with a
characteristic $2\to 2$ cross-section of the \emph{strong} interaction. In the remaining
$3/4$ of the time, the $X$ will decay to a $Z$ or $W$ boson.  Then, leptonic $Z$
decays can be used to discover the $X$ itself.
}
\eitem
In \Sec{constraints} we will show that flavor/$CP$ constraints indeed allow a window for
the case (B).

Note that possibilities (A) and (B) are common to all the 22 models identified in
\Sec{unification}.  For example, in models containing $Q$ instead of $X$, the $\la$
coupling in \Eq{benchmark} should be replaced by $\la_u H u^\C Q + \la_d H^* d^\C Q$,
which would exhibit the same phenomenology as above.  In models containing $U$ or $D$
instead of $X$, the $\la$ coupling is replaced by $\la H U^\C q$ and $\la H^* D^\C q$,
respectively, which is again phenomenologically equivalent.

Actually, the models with $Q$, $U$ and/or $D$ have additional potentially interesting
modes $Q \to Z+t$, $Q \to W+t$, $U \to Z+t$, or $D \to W+t$. While the appearance of
the top constitutes a qualitative difference in the collider phenomenology, a full
analysis for such decay channels is more complicated due to the higher final-state
multiplicity from the top decay. There are reducible background sources which cannot
be simulated reliably at the matrix element level due to the large number of final-state
particles, and even for irreducible backgrounds the issue of combinatoric backgrounds
makes searches more difficult. Some of these problems may be ameliorated if an $\cO(1)$
fraction of tops is produced with large $p_\T$, such that the recently developed methods
of boosted top-tagging~\cite{boostedtop} can be applied. We will leave these more complicated
cases to future work and focus in this paper on the phenomenology of signatures (A) and (B).

\section{Flavor/$CP$ and electroweak constraints}
\Secl{constraints}

In this section, we analyze flavor/$CP$ violations as well as
corrections to precision electroweak observables in the benchmark
model. These corrections arise due to the coupling%
\beq
  \cL \supset \la_{ia} \, H d^\C_i X_a + \cc  \,,
\eql{HdX}
\eeq
where $a=1,2$ and $i=1,2,3$. We will adopt the most conservative
assumption that $\la$ is an ``anarchic'' matrix without any special
texture or alignment:%
\beq
  (\la_{ia})
  =  \lt( \begin{array}{ccc}
             \sim\la & \sim\la \\
             \sim\la & \sim\la \\
             \sim\la & \sim\la
          \end{array}
     \rt).
\eeq
Therefore, the bounds discussed in this section could be relaxed by further model
building or extra assumptions on the structure of $\la$. (For example, \Ref{mirror}
introduces a model with a single $X$ particle (and no $V$) with a similar coupling
selectively to the third generation, and consequently their model is much less
constrained by flavor/$CP$.)

Since the $X$s are heavier than all SM particles and we anticipate a small $\la$,
we integrate out the $X$s and analyze effective operators in powers of $\la/m_X$.
Strictly speaking, the ratio $\avg{H}/m_X$ is not a very small number, so
contributions higher order in $\avg{H}/m_X$ can change our estimates by an
$\cO(1)$ factor. However, our interest is to show that a robust ``Higgs factory''
window can exist for the broad scenario of extending the SM with WIMP DM and
unification, rather than placing precise bounds on this particular benchmark
model.  Therefore, order-of-magnitude estimates suffice for this purpose.

\begin{figure}[t]
\includegraphics[scale=0.4]{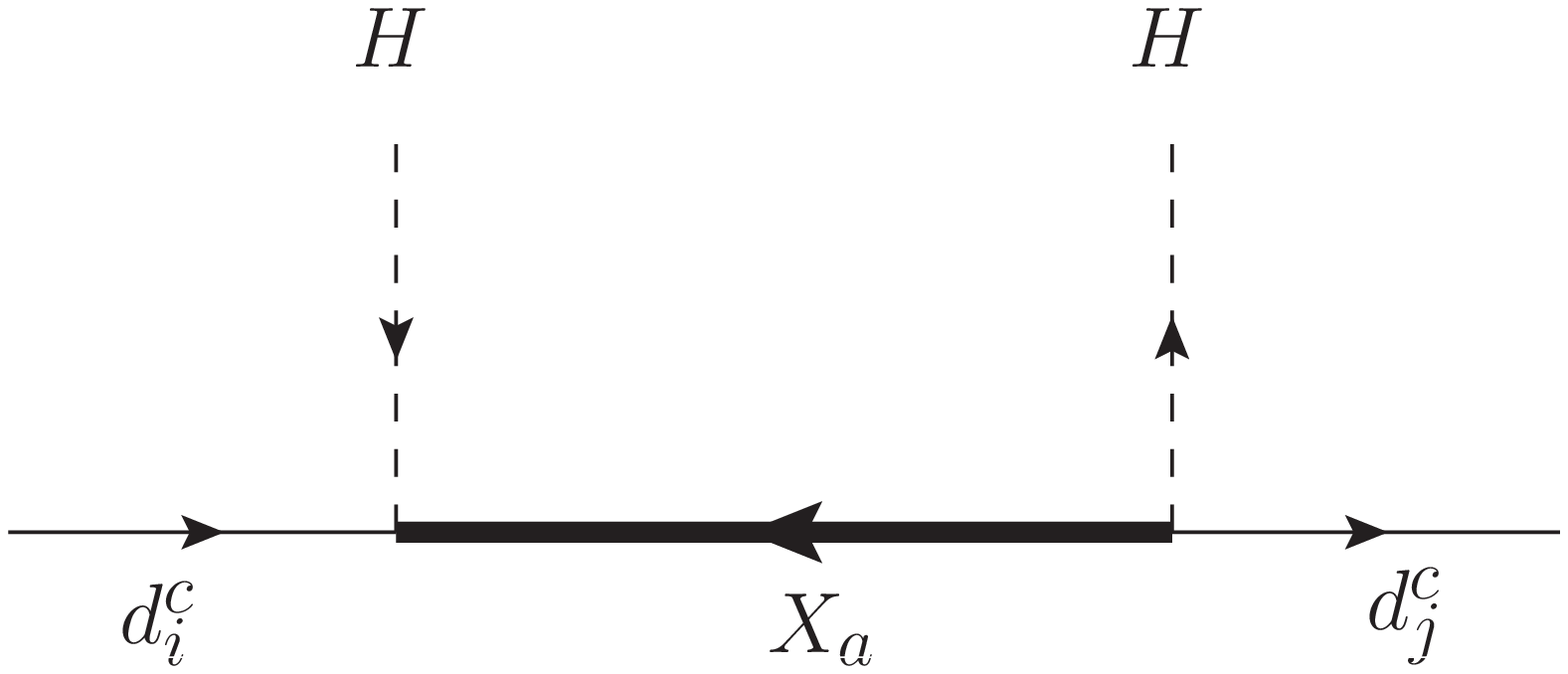}
\caption{\Figl{operator:ddHH}Diagram leading to the operator \eq{tree}.}
\end{figure}
\begin{figure}[t]
\includegraphics[scale=0.4]{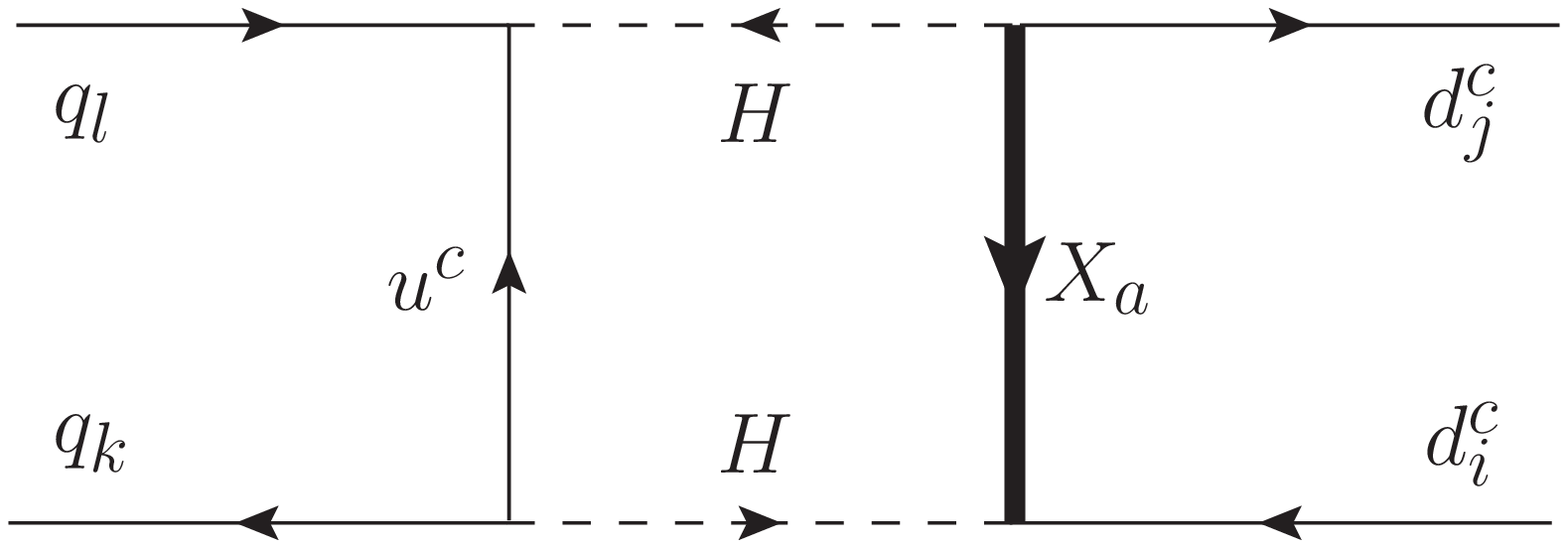}
\caption{\Figl{operator:ddqq}Diagram leading to the four fermion operator \eq{4ferm}.}
\end{figure}

The most stringent bound comes from $K{}^0$-$\overline{K}{}^0$ mixing. The relevant
tree and 1-loop diagrams are shown in Figs.~\fig{operator:ddHH} and \fig{operator:ddqq},
respectively. Upon integrating out $X$ in \Fig{operator:ddHH}, we generate the operator%
\beq
  \sum_{i,j,a} \fr{\la_{ia}^\phtm{\dag} \la_{aj}^\dag}{m_{X_a}^2} \,
  d^{\C}_i \sg^\mu \bar{d}^\C_j \, (H^\dag \DD_\mu H)  \,.
\eql{tree}
\eeq
Below the $Z$ mass, using this operator twice with all the four $H$s put to VEVs would
generate 4-fermion operators with four right-handed down-type quarks with a coefficient
$\sim \la^4 v^2/m_X^4$, which is conservatively $\sim \la^4 /m_X^2$.  In particular, the
imaginary part of the coefficient of $(d^\C \sg_\mu \bar{s}^\C) (d^\C \sg^\mu \bar{s}^\C)$
is constrained and has to be less than  $(10^4\TeV)^{-2}$~\cite{DeltaF=2}.  With our
assumption of anarchic $\la$, we expect an $\cO(1)$ phase in $\la^4$, so we obtain the
bound%
\beq
  \la \lsim 10^{-2} \sqrt{\fr{m_X}{1\TeV}}  \,.
\eql{weaker}
\eeq
Similarly, upon integrating out $X$ in \Fig{operator:ddqq}, we generate the operator%
\beq
  \sim \fr{1}{16\pi^2} \sum_{a,i,j,k,\ell}
       \fr{\la_{ia}^\phtm{\dag} \la_{aj}^\dag (y_u^\dag y_u^\phtm{\dag})_{k\ell}^\phtm{\dag}}
          {m_{X_a}^2}
  (d^\C_i \sg_\mu \bar{d}^\C_j) (\bar{q}_k \bar{\sg}^\mu q_\ell)  \,.
\eql{4ferm}
\eeq
From this, the imaginary part of the coefficient of
$(d^\C \sg_\mu \bar{s}^\C) (\bar{d} \bar{\sg}^\mu s)$
can be estimated to be%
\beq
   \sim \fr{1}{16\pi^2} \fr{\la^2 \tha_\mrm{c}^5}{m_X^2}  \,,
\eeq
where we have used
$(y_u^\dag y_u^\phtm{\dag})_{12}^\phtm{\dag} \sim y_t^2 \tha_\mrm{c}^5 \approx \tha_\mrm{c}^5$
with the Cabbibo angle $\tha_\mrm{c} \approx 1/5$. This coefficient should be less than
$(10^5\TeV)^{-2}$~\cite{DeltaF=2}, which implies%
\beq
  \la \lsim 10^{-2} \fr{m_X}{1~\text{TeV}}  \,.
\eql{flavorbound}
\eeq
The bound \eq{flavorbound} is slightly stronger than the bound \eq{weaker} for
$m_X \lsim 1\TeV$, so we adopt \Eq{flavorbound} as our upper bound on $\la$.

\begin{figure}[t]
\includegraphics[scale=0.4]{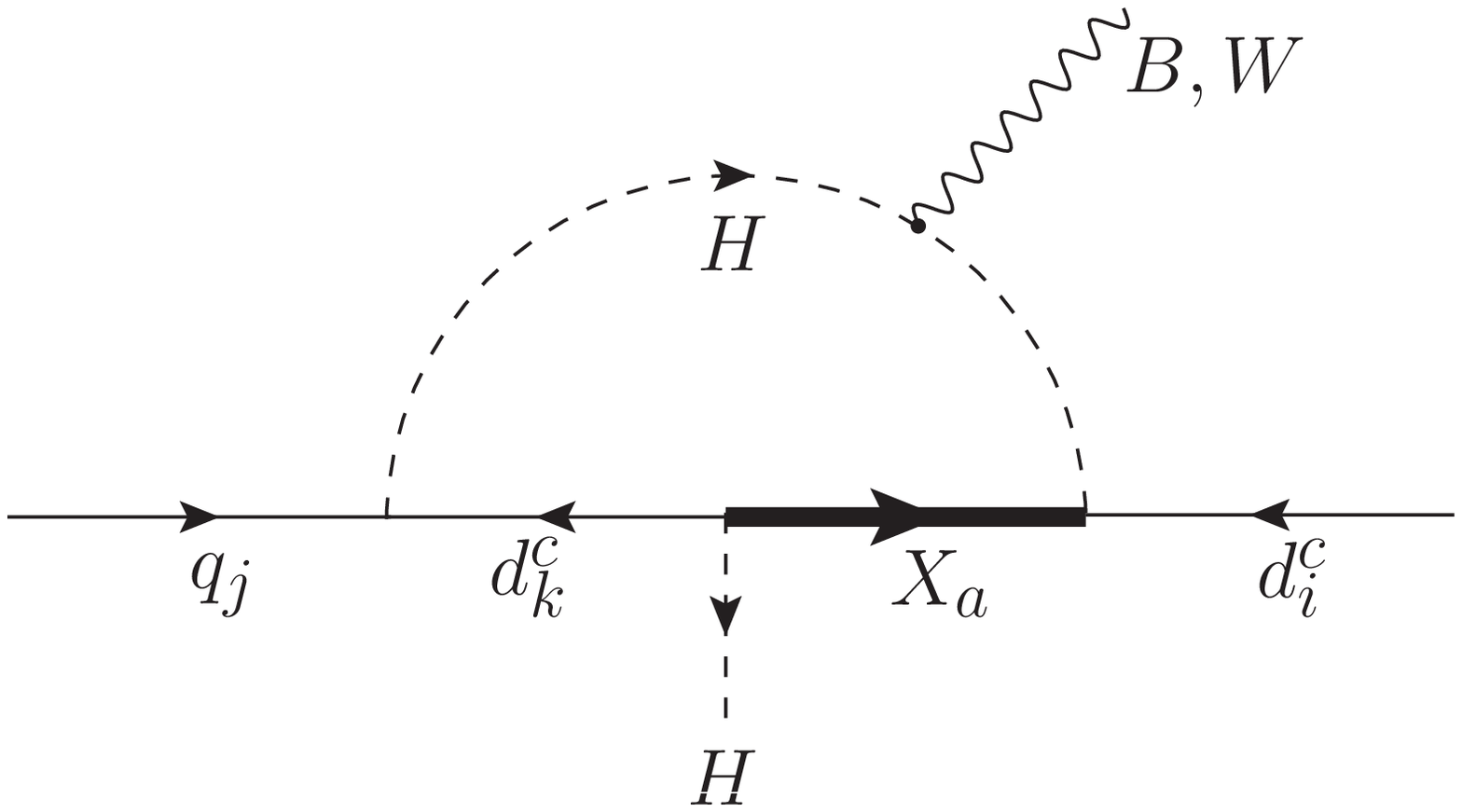}
\caption{\Figl{operator:dipole}One of the diagrams leading to the dipole operator \eq{dip}.}
\end{figure}

No other constrains are as strong as \Eq{flavorbound}.
For example, let us look at the dipole operators generated from diagrams as in
\Fig{operator:dipole}:%
\beq
  \sim \fr{g_F}{16\pi^2} \sum_{a,i,j,k}
       \fr{\la_{ia}^\phtm{\dag} \la_{ak}^\dag (y_d)_{kj}^\phtm{\dag}}{m_{X_a}^2}
       H^\dag d^\C_i \bar{\sg}^{\mu\nu} F_{\mu\nu} q_j  \,,
\eql{dip}
\eeq
where $F=B,W$ denotes the electroweak gauge fields, which in particular contribute
to $b \to s\ga$ after electroweak symmetry breaking.  Since these operators are
suppressed by the small bottom Yukawa coupling $y_b \sim 1/40$, one can think
of them as the $b\to s\ga$ dipole operator in minimal flavor violation~\cite{MFV}
with the scale $\La \sim 4\pi \tha_\mathrm{c} m_X / (\sqrt{e}\, \la)$, where $e$
is the QED coupling.  Then, the bound $\La \sim 10~\text{TeV}$ from
$b\to s\ga$~\cite{MFV, DeltaF=1} implies $\la/m_X \lsim 10^{-1}\text{TeV}^{-1}$,
which is again weaker than \Eq{flavorbound}.

Let us also look at precision electroweak constraints.
First, the operator \eq{tree} modifies $Z \to b\bar{b}$:%
\beq
  \fr{\De g_{Z b \bar{b}}}{g_{Z b \bar{b}}} \sim \fr{\la^2}{m_X^2}  \,.
\eeq
Again, given \Eq{flavorbound}, this is safely below the experimental bound.%
\footnote{\Ref{mirror} discusses a model with a single generation of $X$ (and without $V$),
in which they exploit this shift in $Z \to b\bar{b}$ to improve precision electroweak fits
of the SM.}
Second, \Fig{operator:rho} generates an operator contributing the $\rho$ parameter
\beq
  \sim \fr{\la^4}{16\pi^2 m_X^2} (H^\dag D_\mu H)(H^\dag D^\mu H)  \,.
\eql{rho}
\eeq
The coefficient should be less then $\sim 10^{-3}/v^2$~\cite{EWPT} with $v=174\GeV$,
implying $\la \lsim \sqrt{m_X/(1\TeV)}$, which again is much weaker than
\Eq{flavorbound}.

\begin{figure}[t]
\includegraphics[scale=0.4]{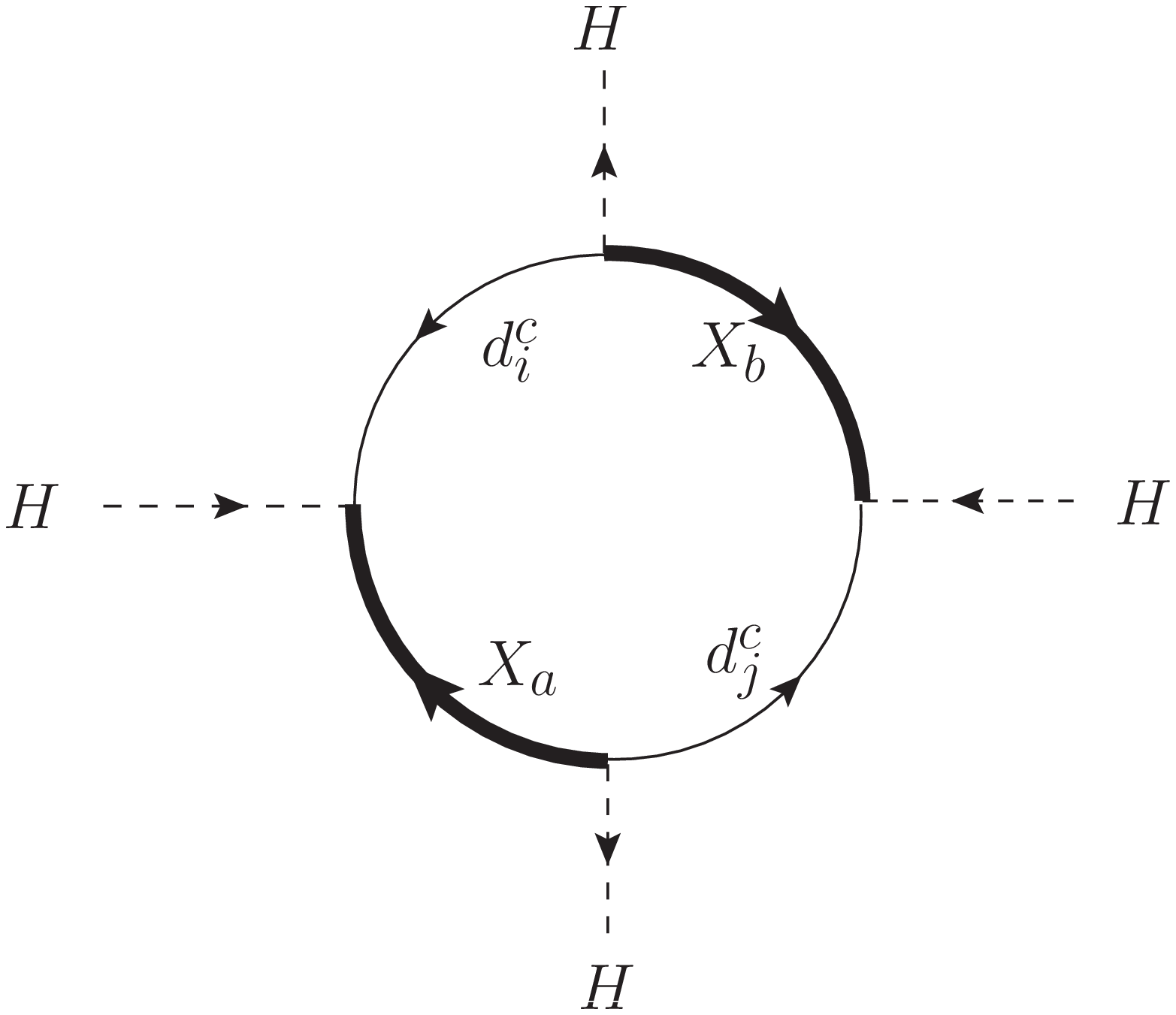}
\caption{\Figl{operator:rho}Diagram contributing to the $\rho$ parameter \eq{rho}.}
\end{figure}

Given the bound \eq{flavorbound}, there is clearly a robust ``Higgs factory'' window where
the $X$s decay promptly. Keeping only the $X$ mass for simplicity, the decay rate of an $X$
via coupling \eq{HdX} is given by $\Ga_X \sim \la^2 m_X/16\pi$. Therefore, demanding
$\Ga_X^{-1} \lsim 10^{-12}\s$ (i.e.~the decay length shorter than $\cO(0.1)\mm$),
we obtain the Higgs-factory window
\beq
   10^{-7} \lt( \frac{1~\text{TeV}}{m_X} \rt)^{\!\!\fr12}
   \lsim
   \la
   \lsim 10^{-2} \frac{m_X}{1~\text{TeV}}  \,.\eql{Hfact}
\eeq

Below the Higgs-factory window, there is a two or three orders-of-magnitude window where an
$X$ decays with a displaced vertex or in the LHC detector but with a macroscopic decay length.
At a hadron collider, events where new physics only manifests itself at a macroscopic distance
from the interaction point but still within the volume of the detector are challenging. Such
signatures also appear in supersymmetric theories~\cite{GM,slowsusy}, hidden valley
models~\cite{HV}, quirk models~\cite{quirk}, and vectorlike confinement models~\cite{VC}.
As discussed in section 16 of \Ref{Brooijmans:2010tn} (and references therein), when particles
with macroscopic decay lengths are produced, trigger efficiencies can become a concern. Recently,
the prospects of a similar final state from late-decaying neutralinos in a supersymmetric model
have been carefully studied in \Ref{Meade:2010ji}, with encouraging results. While the case of
macroscopic decay lengths is interesting, the subtleties involved require a study with a more
sophisticated detector simulation than {\tt PGS}~\cite{PGS} (which we use in our analysis).
In addition, it is possible for
R-hadrons to stop in the detector~\cite{stopping}. These are important questions worth
investigating in detail, which we leave to future work.

\section{Collider Phenomenology}
\Secl{Pheno}
In this section we will investigate in detail the two characteristic collider signatures
of our scenario using the benchmark model. When the $X$s are collider stable, we will show
that the discovery of $X$ is possible up to $m_{X}=650\GeV$ in the early $7$-TeV run with
$1\fb^{-1}$ of data, and well past $1\TeV$ with the $14$-TeV running. When the $X$s decay
promptly, we will concentrate on the discovery of the $X$ and the Higgs bosons from $X$
decays for early 14-TeV running ($10\fb^{-1}$). In order to keep the analysis simple, we
will restrict ourselves to the case of a moderately heavy Higgs ($m_{h}=200\GeV$), but we
expect the Higgs discovery potential to be similarly enhanced for a light Higgs as well.

Even though the benchmark model contains two generations of $X$, we have no reason to
expect that they would be exactly degenerate in mass. Since the production cross-section
will be dominated by the lighter $X$, we will base our collider analysis on one generation of $X$
only. This is the most conservative choice; if the $X$s do happen to be nearly degenerate,
this would significantly enhance the results below.

\subsection{R-hadron Signals at the LHC}
\Secl{Rhadron}

When the coupling $\la$ is sufficiently small, the $X$ does not decay within the detector.
The signature therefore is that of an ``R-hadron'', that is, the $X$ hadronizes with light
colored degrees of freedom and the color-neutral bound state behaves as a (possibly charged)
stable massive particle. Before we present a quantitative analysis, let us dwell on a few
qualitative features of this signature.

Firstly, note that even when $\la$ is very small, $X_{-4/3}$ still decays to $X_{-1/3}$
via the weak interactions, and in terms of collider signatures, the production of $X_{-4/3}$
is indistinguishable from that of $X_{-1/3}$ because the decay products are virtually
unobservable. As worked out in detail in \Ref{minimalDM}, $X_{-4/3}$ is expected to be
heavier than $X_{-1/3}$ by only $\Delta m_X=0.60\GeV$. The dominant decay mode is
$X_{-4/3}\to X_{-1/3}+\pi^-$ through an off-shell $W^-$, with a partial width%
\beq
  \Gamma^{-1}
  = 1.3\mm \left( \frac{0.60\GeV}{\Delta m_X} \right)^{\!\!3}
    \sqrt{ \frac{1- \frac{m_{\pi}^2}{(0.60\GeV)^2} }
                {1- \frac{m_{\pi}^2}{\Delta m_X^2} }
         }  \,.
\eeq
The smallness of the mass gap makes the $\pi$ very soft and thus unobservable at the LHC,
and other subdominant decay modes have the same problem. Therefore, in analyzing the
discovery potential or checking existing bounds, the production cross-section of $X_{-4/3}$
should be added to that of $X_{-1/3}$.

The charge of an R-hadron is crucial for prospects of observing it. In particular, the
most effective way to trigger on a stable charged massive particle is via the muon
system~\cite{RhadronATLAS}. The bound states of an $X_{-1/3}$ can be mesons ($X\bar{q}$)
or baryons ($Xqq$).  The physics of these bound states can be understood by regarding the
$X_{-1/3}$ as a heavy version of the $b$ quark, which is already much heavier than
$\La_\text{QCD}$. The lightest $B$ mesons are $B^0$ and $B^\pm$, with only a few-hundred-keV
mass splitting, while the lightest $B$ baryon $\La_b^0$ is heavier than the $B^{0,\pm}$ by
$340\MeV$.  Therefore, we expect that the lightest $X$-meson should be lighter than the
lightest $X$-baryon also by $\sim 340\MeV$, with a few-hundred-keV mass splitting between
the neutral and charged $X$-mesons. Since the splitting between the lightest $X$-meson and
$X$-baryon is on the order of $\La_\text{QCD}$ itself, we expect that an $X$ should
preferentially hadronize into an $X$-meson, which can be either charged or neutral with
50\% probability because their mass difference is tiny.%
\footnote{The few-hundred-keV mass difference between the $(X_{-1/3} \bar{d})$ meson
and the $(X_{-1/3} \bar{u})$ meson might allow one to decay to the other via the weak
interaction, if the mass difference is larger than the electron mass.  But such a
decay would occur with an extremely long lifetime (like the $\beta$-decay of the
neutron), so it can be ignored on collider time scales.}

In order to estimate the trigger efficiencies, we will
use the following assumptions in the rest of our analysis:
\begin{itemize}
\item{There is a 50\% chance that an R-hadron is charged when
produced at the primary interaction.
}
\item{This charge is retained until the calorimeter is reached.
}
\item{One or more charge exchange interactions take place in
the calorimeter, randomizing the charge of the R-hadron such
that there is a 50\% chance that it reaches the muon chamber
as a charged particle.%
\footnote{In addition to conversion between the charged and neutral $X$-mesons, for
which there is a tiny energetic cost of a few hundred keV, there is also a process where
an $X$-meson scatters into an $X$-baryon in the calorimeter~\cite{Kraan}, which, however,
requires an energy of at least $\sim 340\MeV$.  Using the analogy with the $b$ system, the
lightest $X$-baryon (analogous to $\La_b^0$) should be neutral, while the lightest charged
$X$-baryon (analogous to $\Sg_b^\pm$) should be heavier by $\sim 190\MeV$. So, a charged
$X$-baryon, even if produced, would promptly decay to a neutral $X$-baryon (by emitting a
pion) which would then not be caught by the muon chamber, potentially hurting our R-hadron
signal. The question of how frequently this meson-to-baryon conversion occurs is highly
nontrivial and beyond the scope of this paper.  However, note that a meson-to-baryon
conversion would not occur to the $\wba{X}$ due to the lack of anti-nucleons in the
calorimeter. Therefore, even in the worst case where we always lose the R-hadrons from
the $X$, we still have those from the $\wba{X}$, so the effective cross-section would
be roughly halved, which would correspond a small shift ($\sim \cO(10)\GeV$) in the
mass scale.}
}
\end{itemize}
One of the requirements for triggering is that the particle reaches the muon system with
nonzero charge. Most experimental searches for massive stable particles also use as a
selection criterion that there should be a charged track in the inner part of the detector
that matches the hit in the muon chamber, even if this is not required for triggering.
Therefore we will also adopt this as one of our event selection criteria.

Finally, an important kinematic variable, especially at the LHC (where the detectors
are physically larger and the time between bunch crossings is short), is the ``time-lag''.
This is defined as how much later the massive R-hadron ($\beta < 1$) arrives at the muon
chamber compared to a relativistic particle (e.g.~a muon). To be conservative, we use the
physical dimensions of the ATLAS detector (which is larger than CMS) as described in
\Ref{RhadronATLAS}. Specifically, we differentiate the barrel region ($|\eta|<1.4$) from
the end-cap region ($1.4<|\eta|<2.5$). For the barrel region we calculate the time to get
to a radius of $7.5~{\rm meters}$ from the interaction point, and for the end-cap we
calculate the time to get to $|z|=14.5~{\rm meters}$. For the LHC search, we will use
as one of the event selection criteria that at least one of the R-hadrons reaches the
muon chamber before relativistic particles from the next bunch crossing, i.e., with a
time-lag of less than $25\ns$.

\begin{figure}
\includegraphics{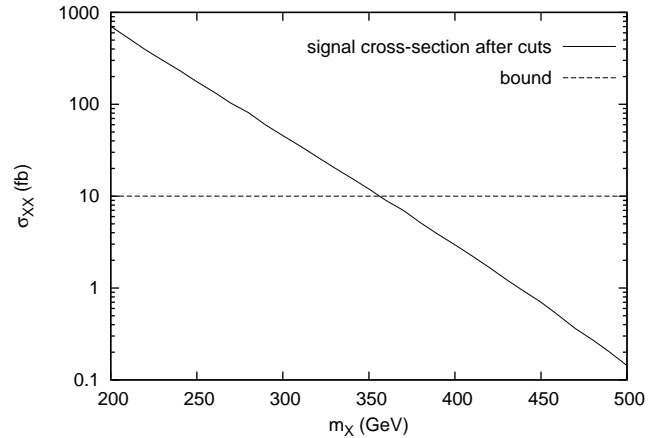}
\caption{\Figl{tevbound}The cross-section for R-hadron production at the Tevatron
after all selection cuts. The bound from the CHAMP search is included for comparison.}
\end{figure}

Before we investigate the discovery potential of $X$ at the early LHC, let us address
constraints on R-hadron production from the Tevatron~\cite{TeVCHAMPsearch, Abazov:2008qu}.
We will use the event selection criteria described in \Ref{TeVCHAMPsearch} in order to
estimate acceptance and trigger rates in the benchmark model. In particular we demand
that events contain at least one particle that has $|\eta|<0.7$, $p_\T>40\GeV$ and
$0.4 < \beta < 0.9$, leaves a track, and is charged when it arrives at the muon system.
For a single R-hadron satisfying these cuts, our above assumptions on the charges of R-hadrons
imply a 25\% probability for being detected.  However, when both of the pair-produced R-hadrons
are within acceptance and charged throughout the detector (a $1/16$ probability), we need to
correct for the fact that the reconstruction and trigger efficiency used in the analysis of
\Ref{TeVCHAMPsearch} applies to a single R-hadron. Therefore the ``effective cross section''
for such events needs to be multiplied by a correction factor $\xi$ before the comparison
with the bound of \Ref{TeVCHAMPsearch}.
This correction factor is given by $\xi=(1-(1-\ep)^2)/\ep$, where $\ep=0.533$
is the reconstruction efficiency for a single R-hadron~\cite{todd}. We
use {\tt CalcHEP 2.5.4}~\cite{calchep} with CTEQ6 parton distribution
functions~\cite{CTEQ6} to simulate the parton level process. The
effective cross-section after the selection cuts is plotted in
\Fig{tevbound} against the bound of \Ref{TeVCHAMPsearch}. We
conclude that $m_{X}>360\GeV$ is not excluded.

\begin{figure}
\includegraphics{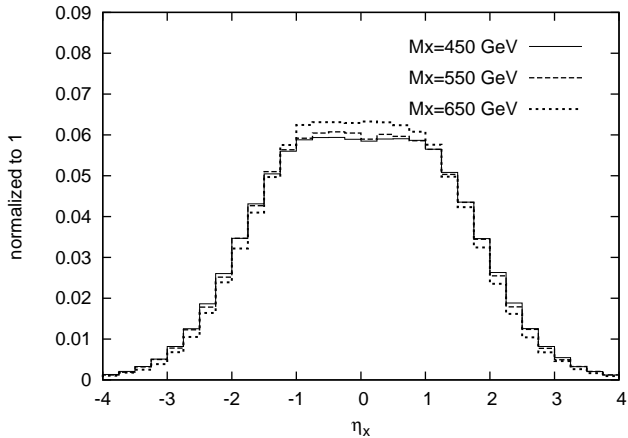}
\caption{\Figl{eta}The rapidity distribution of the R-hadrons at the LHC ($7\TeV$) for the
three mass points.}
\end{figure}

We now turn to production of R-hadrons at the early LHC, i.e., for the 7-TeV
running with $1\fb^{-1}$ of integrated luminosity.  We choose to look at mass
points $m_{X}=450\GeV$, $550\GeV$ and $650\GeV$. In \Fig{eta}, we plot the rapidity
distribution of the R-hadrons for each mass point to show that R-hadron production
is dominantly central. As event selection criteria, we impose that at least one
R-hadron must reach the muon detector ($|\eta|<2.5$) with $p_\T>30\GeV$ and a time
lag of less than $25\ns$. We further demand that the R-hadron in question leaves a
track and is charged when it gets to the muon chamber (a 25\% probability per R-hadron
as before). In events where both R-hadrons reach the muon chamber with $p_\T>30\GeV$,
we plot the time-lag of the earlier (later) R-hadron in \Fig{tof1} (\Fig{tof2}).
Folding in the time-lag cut and the probability of being charged (but leaving out
reconstruction efficiencies, which are unknown at this time), we plot the
effective cross-section after all selection cuts in \Fig{mp1}. Note that the efficiency of the selection cuts has a slightly decreasing trend at higher mass, because the production occurs closer to threshold and fewer events satisfy the time-lag cut. Requiring at least
10 events for discovery, we see that $X$ masses of up to $650\GeV$ should be within
reach with $1\fb^{-1}$ of data from the 7-TeV run.

\begin{figure}
\includegraphics{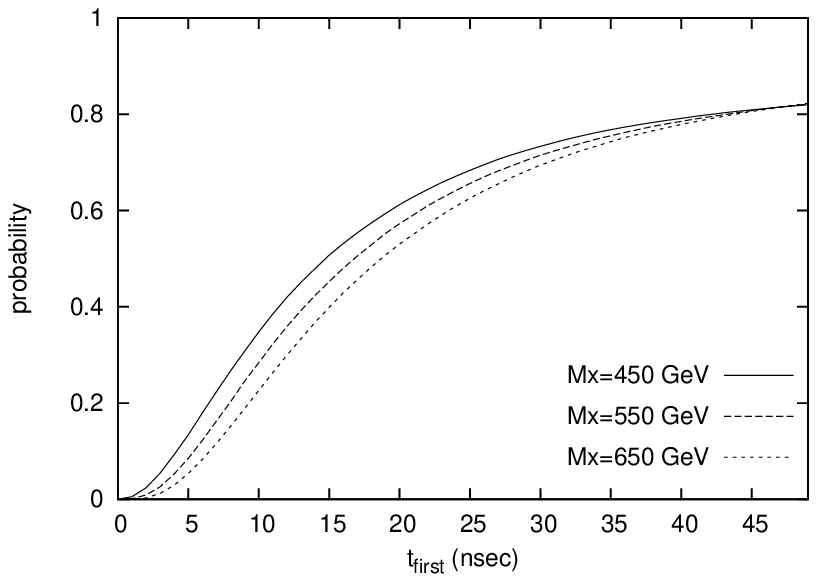}
\caption{\Figl{tof1}Time-lag distribution of the first R-hadron to arrive in the muon system
at the LHC ($7\TeV$).}
\end{figure}
\begin{figure}
\includegraphics{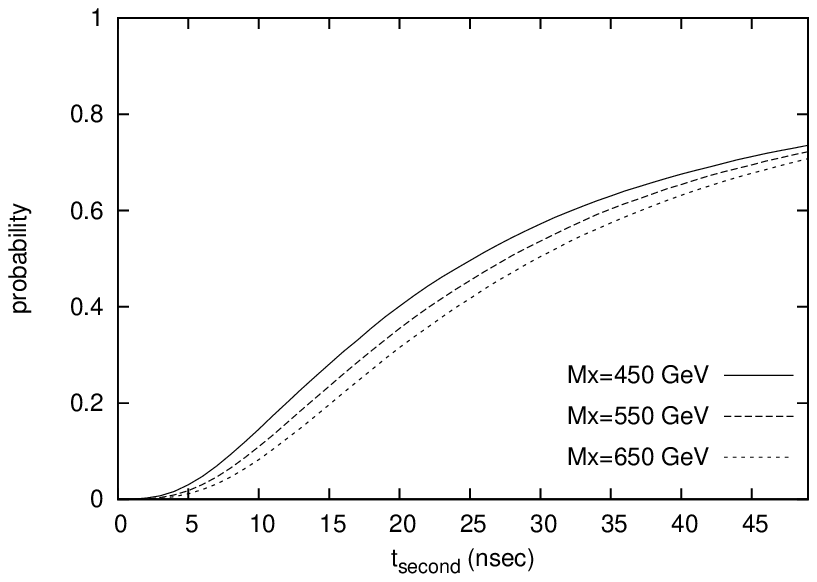}
\caption{\Figl{tof2}Time-lag distribution of the second R-hadron to arrive in the muon system
at the LHC ($7\TeV$).}
\end{figure}
\begin{figure}
\includegraphics{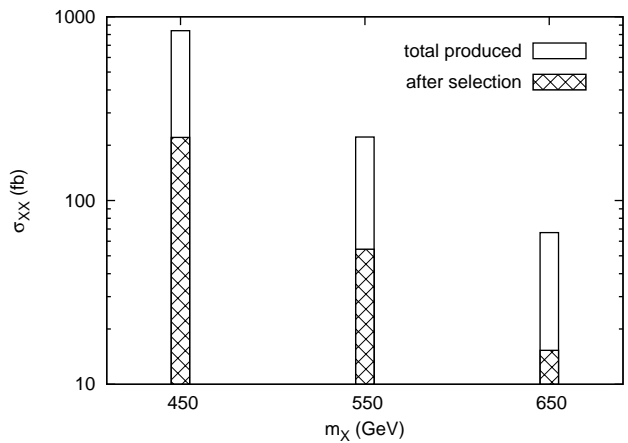}
\caption{\Figl{mp1}Total production cross-section of R-hadron production and effective
cross-section after selection cuts at the LHC ($7\TeV$) for the three mass
points.}
\end{figure}
\begin{figure}
\includegraphics{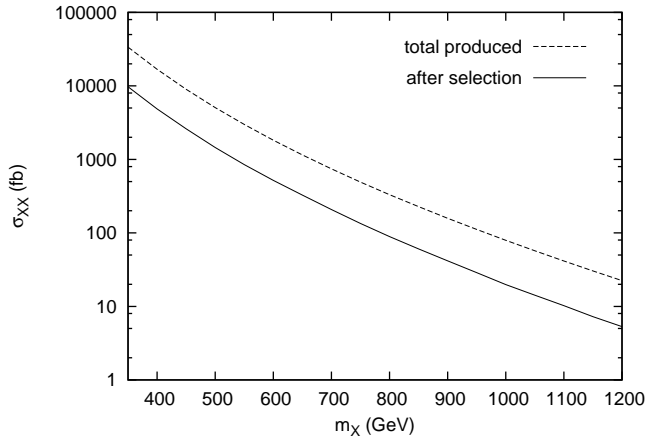}
\caption{\Figl{mp2}Effective R-hadron production cross-section after selection cuts at the
LHC ($14\TeV$).}
\end{figure}

Using the same selection criteria for the 14-TeV running, we also plot in \Fig{mp2}
the effective cross-section after cuts in that case. We see that even with early 14-TeV
data (e.g.~$\approx 10\fb^{-1}$), $X$ masses well past $1\TeV$ should be within reach.

\subsection{The LHC as a Higgs factory}
\Secl{higgsfac}
For $\lambda$ in the range given by \Eq{Hfact}, the $X$ decays promptly. For the $X_{-1/3}$,
the dominant decay modes are $Z+j$ and $h+j$. Since production proceeds through QCD, this
gives rise to a ``Higgs factory'' if $m_{X}$ is not too large. The $X_{-4/3}$ decays to $W+j$
and therefore gives no additional contribution to Higgs production. For the rest of this
section, we will be interested in the production of $X_{-1/3}$ only. We will show that there
is a window in $m_{X}$, where with $10\fb^{-1}$ of 14-TeV running, both the $X$ and the Higgs
can be discovered.

We will first dwell on the discovery of the $X$ when both of the pair-produced $X$s decay to
$Z+j$. For the purposes of this paper, we will restrict ourselves to looking at leptonic decay
modes of the $Z$ as it has considerably less background, but a full collider study can combine
various channels and extend the reach in $m_{X}$.

We will then use the value of $m_{X}$ extracted from this analysis in order to discover the
Higgs in the events where one of the $X$s decays to $Z+j$ and the other to $h+j$. We will focus
on a scenario with a moderately heavy Higgs, with a dominant decay mode to $W^{+}W^{-}$, although
discovery of a light Higgs through $X$ production should be competitive with Higgs production
from the SM as well (possibly utilizing the recent search methods involving boosted final
states~\cite{boostedHiggs, boostedstuff}).  Once again, we will limit ourselves to leptonic
decays of the $Z$ as well as the $W$s, but in a more detailed analysis, several channels can
be combined to extend the discovery reach.

Let us start with the Tevatron bounds on $m_{X}$.
The strongest constraint arises
from a recent analysis of $WZ$ production~\cite{WZlimits}, searching for events with exactly three final state leptons. In our benchmark model, events with pair-produced $X$ decaying to $Z$, $h$ and jets, followed by $h\to WW$ can be picked up by this search, as well as $ZZ$ plus jet final states when both $Z$ decay leptonically but one lepton fails to be identified.  We implemented the cuts described in \Ref{WZlimits}, which include requiring the presence of an $e^{+} e^{-}$ or $\mu^{+} \mu^{-}$ pair with invariant mass in the interval $[86~{\rm GeV},106~{\rm GeV}]$ as well as $p_{T,\ell_{1}}>20~{\rm GeV}$, $p_{T,(\ell_{2},\ell_{3})}>10~{\rm GeV}$ and $\slashed{E}_T>25~{\rm GeV}$. We find that for $m_{X}>300\GeV$, the number of
events from $X$ production after the selection cuts falls within one standard deviation of the SM expectation.
The $WW$ final state poses a weaker constraint, because in this final state a veto on
hard jets is imposed in order to reduce the $t\bar{t}$ background~\cite{WWlimits}. For
$m_{X}>300\GeV$ the $X$-production cross-section is only a fraction of the $t\bar{t}$
cross-section, so there is no additional constraint from searches for $t\bar{t}$
either. Finally, for the same range in $m_X$, the Higgs production from $X$ decays is
significantly below the SM Higgs production cross-section for the same $m_h$, so Higgs
searches also do not give additional constraints. In order to stay consistent with
these constraints, we will concentrate on the rest of this section on two mass
points, $m_X=300\GeV$ and $m_X=550\GeV$, both with $m_h = 200\GeV$.

In our analysis, we generate  $3\times10^{5}$ parton level events (at 14-TeV LHC running) for each value of $m_X$ using the user-mode
of {\tt MadGraph}~\cite{MG} and CTEQ6 parton distribution functions~\cite{CTEQ6}. We decay the
$X$, $h$, $W$ and $Z$ particles using {\tt BRIDGE}~\cite{BRIDGE} in order for the angular
distributions of the final-state leptons and jets to be accurate.  In the signal sample we allow all possible decays of $X$, $h$, $W$ and $Z$
in order to take full account of the combinatoric background, while in the background samples we force leptonic decays (including $\tau$'s) since hadronically decaying background events will not pass our selection cuts. The hadronization and
detector simulation are done with the {\tt PYTHIA}~\cite{PYTHIA} and {\tt PGS}~\cite{PGS}
interface in {\tt MadGraph}, with the default CMS parameter set. This parameter set uses a tracker and muon system $\eta$ coverage up to 2.4, and a minimum lepton $p_{T}$ of $5~{\rm GeV}$. Cone jets with $\Delta R=0.5$ are used, note however that our analysis is inclusive and we do not place any cuts on jets. The {\tt PGS} default algorithms are used for lepton isolation as well as other details of detector simulation. As the background for
the $X$ search we generated a matched  $ZZ+\text{jets}$ sample containing 53494 events after matching, using the MLM matching in
{\tt MadGraph}, and the same tools as were used for the signal. For the Higgs search we
generated a matched $t\bar{t}Z+\text{jet}$ sample with 36120 events after matching in the same way.

\subsubsection{Discovering the $X$.}

In order to discover the $X$, we focus on the $ZZ$ final state, where both $Z$s decay
leptonically. For SM backgrounds, we have
generated an MLM-matched sample with $ZZ$, $ZZ+j$, and $ZZ+2j$. As our event selection
criteria, we demand that an event contains two (distinct) $Z$ candidates, where a $Z$
candidate is defined as an $e^{+}e^{-}$ or a $\mu^{+}\mu^{-}$ pair with an invariant mass
within $5\GeV$ of $m_{Z}$. We then pair the two $Z$ candidates with the two hardest jets in
the event (which in the case of signal are expected to come from the partons of the $X$ decays)
and retain the pairing where the $Zj$ pair masses are closer to each other. We then plot the
average pair mass $m_{Zj}$, where the peak from the $X$ is clearly visible and distinguishable
from the
$ZZ+\text{jets}$ background.  For $m_X=300 \GeV$, where the branching fraction $X\rightarrow Zj$ is 0.76, the cross-section
before (after) selection is 36.8 pb (50.5 fb). Similarly, before (after) selection we obtained 1.43 pb (1.32 fb) for $m_X=550 \GeV$, where the branching fraction $X\rightarrow Zj$ is 0.57. Leptonic branching fractions and selection cuts reduce the SM background cross section from 11.4 pb down to 5.1 fb. The results for the two mass points and for $10\fb^{-1}$ are plotted in \Fig{Xrec300} and \Fig{Xrec550}. Since the background peaks towards low values of $m_{Zj}$, a cut on $m_{Zj}$ can improve signal significance. For the $m_X=550 \GeV$ case we use $m_{Zj}\geq 430 \GeV$ as a selection cut. While a similar cut can also be used for the $m_X=300 \GeV$ case, signal is already much larger than background in this case and therefore a cut on $m_{Zj}$ is not essential. With this additional cut, the signal cross section becomes 0.85 fb while background is reduced to 0.074 fb. With an integrated luminosity of $10~{\rm fb}^{-1}$, this translates to an average 9.2 events with a background expectation of 0.74 events. Using Poisson statistics, this corresponds to a probability of $9.4\times10^{-8}$, equivalent to more than a $5\sigma$ upward fluctuation in a Gaussian distribution. We conclude that $m_{X}$ up to $\approx 550\GeV$ is discoverable with $10\fb^{-1}$ at $\sqrt{s}=14\TeV$.

\begin{figure}
\includegraphics{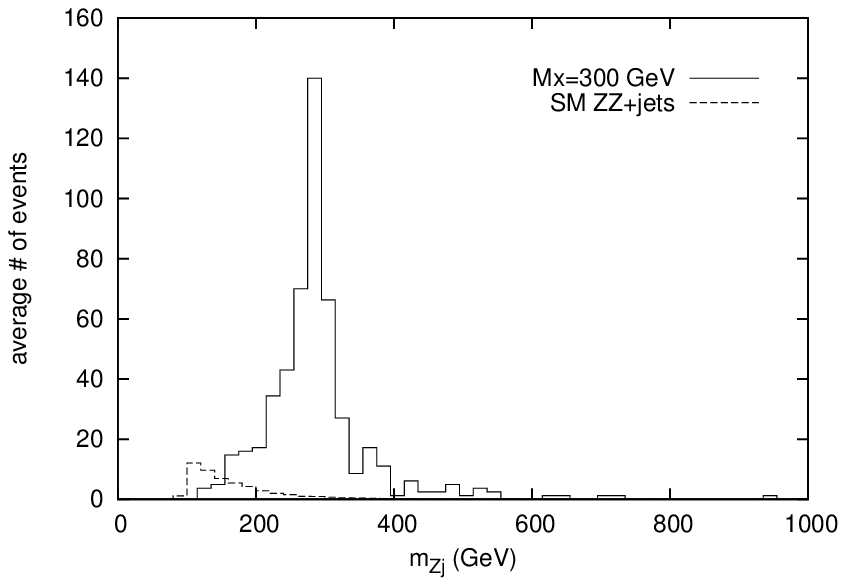}
\caption{\Figl{Xrec300}Distribution of the average $Z$+jet pair mass for signal
($m_X=300\GeV$) and background in $10\fb^{-1}$ of LHC data ($14\TeV$).}
\end{figure}
\begin{figure}
\includegraphics{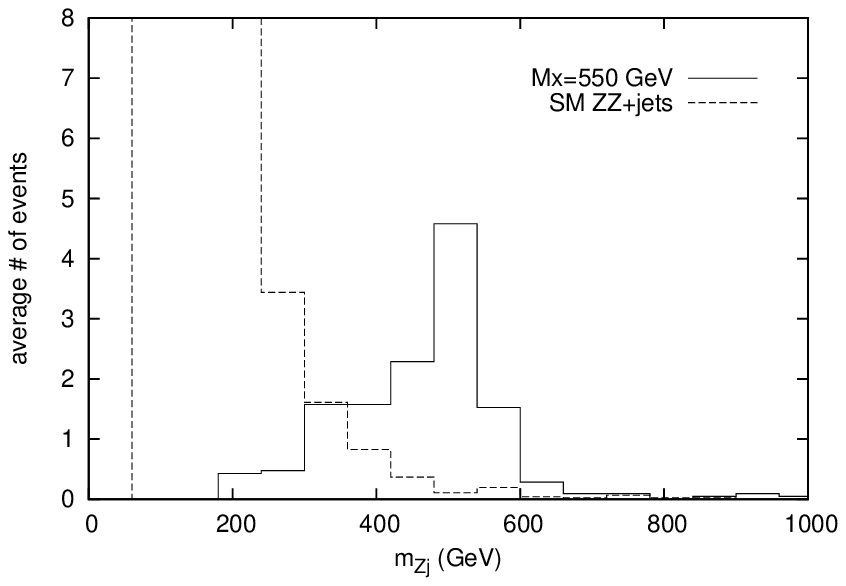}
\caption{\Figl{Xrec550}Distribution of the average $Z$+jet pair mass for signal
($m_X=550\GeV$) and background in $10\fb^{-1}$ of LHC data ($14\TeV$).}
\end{figure}
%

\subsubsection{Discovering the Higgs}

In order to discover the Higgs, we focus on the $Zh$ final state, where the $Z$ as well as
the $W$s from the Higgs decay to leptons. More concretely, our event selection criteria are:%
\begin{itemize}
\item{The event contains two positively and two negatively charged leptons, with
exactly one $Z$ candidate (as defined above).
}
\item{The $Z$ candidate and (at least) one of the two hardest jets has an invariant mass
within $90\GeV$ of $m_{X}$ as determined in the previous section.
}
\end{itemize}
For $m_X=300 \GeV$ the selection reduced the signal cross-section from 36.8 pb to 22.2 fb, for  $m_X=550 \GeV$ from 1.43 pb to 0.87 fb.
With the first selection criterion, the dominant background is $t\bar{t}Z+\text{jet(s)}$.
At parton level, we generated an MLM-matched sample with up to one extra parton,
i.e., $t\bar{t}Z$ and  $t\bar{t}Z+j$.  The second selection criterion, which uses the value for $m_X$ obtained with the search strategy described in the previous subsection, then further reduces the background such that we are essentially left with a pure signal sample. Leptonic branching fractions and the selection cuts reduce the background cross-section from 0.61 pb to 0.40 fb (0.083 fb), applying the two selection criteria for $m_X=300 \GeV$ ($m_X=550 \GeV$). For both mass points, this corresponds to discovery level statistical significance. Using Poisson statistics in the heavy mass case, the probability for a background fluctuation to mimic the signal is $2.4\times 10^{-7}$, equivalent to more than $5\sigma$ in a gaussian distribution.

We then identify the two leptons which do not belong to the $Z$ candidate, and form the
transverse mass variable $M_{\T,WW}$ as follows:%
\beq
  \hspace{-1.5ex}
  M_{\T,WW}^2
  = (E_{\T, l^+ l^-} + E_{\T, \nu \bar{\nu}})^2
   -(\vec{p}_{\T, l^+ l^-} + \vec{p}_{\T,\text{miss}})^2,
\eeq
where%
\beq
  E_{\T,l^+l^-}^2        &=& p_{\T,l^+l^-}^2      + m_{l^+l^-}^2  \,, \nn\\
  E_{\T,\nu \bar{\nu}}^2 &=& p_{\T,\text{miss}}^2 + m_{l^+l^-}^2  \,.
\eeq
Note that $E_{\T,\nu\bar{\nu}}$ is only an approximation to the true transverse energy
of the neutrino system%
\beq
  E_{\T,\nu \bar{\nu}, \text{true}}^2
  = p_{\T,\nu \bar{\nu}}^2 + m_{\nu\bar{\nu}}^2.
\eeq
Although this approximation only becomes exact when the $W$ are produced at threshold,
for a 200-GeV Higgs the $M_{\T,WW}$ distribution still peaks near the Higgs mass. We
plot the results for the $m_X=300\GeV$ in \Fig{MTWW1} and for $m_X=550\GeV$ in
\Fig{MTWW2}.

\begin{figure}
\includegraphics{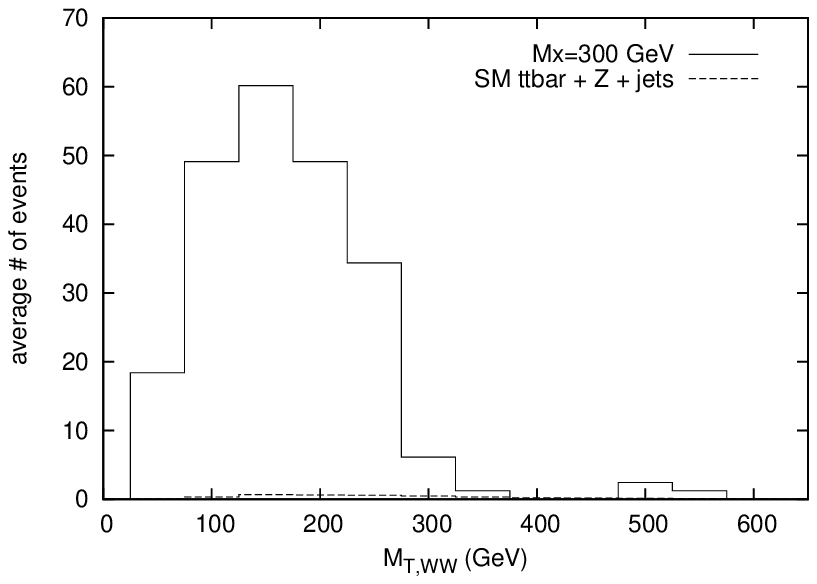}
\caption{\Figl{MTWW1}Distribution of $M_{\T,WW}$ for signal ($m_X =300\GeV$, $m_h=200\GeV$)
and background with a luminosity of $10\fb^{-1}$.}
\end{figure}
\begin{figure}
\includegraphics{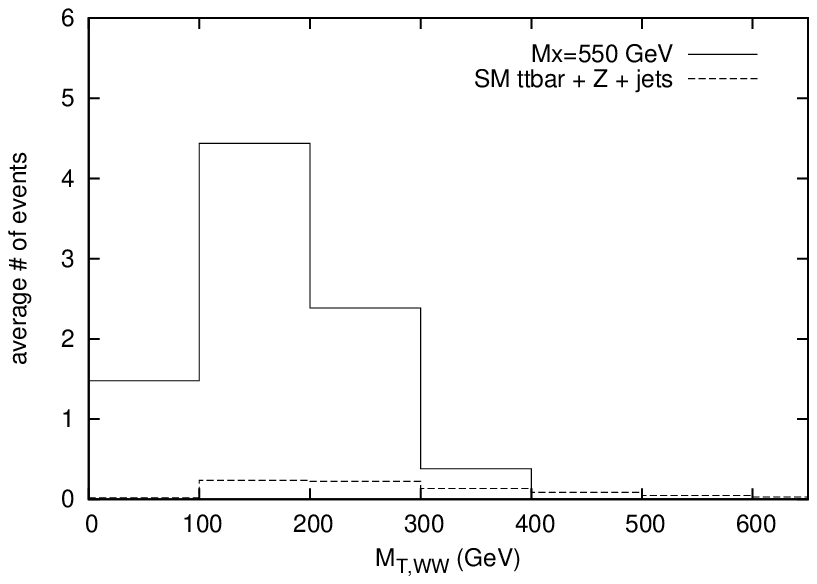}
\caption{\Figl{MTWW2}Distribution of $M_{\T,WW}$ for signal ($m_X =550\GeV$, $m_h=200\GeV$)
and background with a luminosity of $10\fb^{-1}$.}
\end{figure}
%

\section{Concluding remarks}
\Secl{conc}
The minimal way to incorporate a WIMP DM candidate is as the neutral component
of an electroweak triplet with zero hypercharge. We have looked for possible extensions
of the SM that contain such a triplet as well as additional matter fields as necessary for
gauge coupling unification. We have identified the characteristic features of such models.
New colored particles at the TeV scale are ubiquitous, which can be produced at the LHC. The
colored particles allow couplings such that they dominantly decay to a Higgs and a jet, within
or outside the LHC detector depending on the size of the couplings.  The former possibility
gives rise to a new channel for Higgs production, while the latter leads to spectacular R-hadron
signals. In order to study these interesting characteristic collider signatures of WIMP DM and
unification, we have chosen the model with the simplest matter content as a benchmark. The
benchmark model contains two generations of an $\SUL$-doublet color-triplet particle $X$ that
decay via Yukawa-type couplings to the SM. In particular, the final states contain $W$s, $Z$s
and Higgses, as well as down-type quarks.

We have then investigated the constraints from flavor bounds on the size of the Yukawa-type
coupling that leads to the $X$ decay. We showed that there is a range where the $X$ can decay
promptly, or can be long-lived (stable on collider time scales). We have explored each of
these possibilities in turn, showing that in the case of the long-lived $X$, R-hadrons can
be discovered at the early LHC ($7\TeV$) up to $m_X=650\GeV$, and past 1~TeV with a 14-TeV
running.

In the case where the $X$ particles decay promptly, a large number of Higgs bosons are
produced through the decays, which, depending on the $X$ mass, can be discovered with
less luminosity than would be possible from SM Higgs production. We have shown that,
for $m_X \lsim 550\GeV$ and with $10\fb^{-1}$ of data at the LHC, we can discover the $X$ itself
in the leptonic $ZZ$ final state as well as the Higgs bosons from $X$ decays in the leptonic
$WW$ final state for a benchmark Higgs mass $m_h=200\GeV$.

While in this paper our goal was to focus only on signals that are clean and have little
background, these studies can be significantly expanded in a more dedicated collider search.
In particular, semi-leptonic decay channels can be combined with the fully leptonic ones to
increase the reach. The discovery potential of a light Higgs should be enhanced as well, especially
utilizing the recently introduced search methods relying on boosted final states. Finally,
models other than the benchmark model we have chosen can be studied for qualitatively different
final states.  For example, while final states with up-type quarks (in particular the top quark)
are rare in the benchmark model, other models can give rise to a large number of tops produced
from the decays of the new physics (in addition to the two signals we discussed in this paper).
Another interesting problem is to study the case where $X$ decays within the detector with a
displaced vertex or with a macroscopic length. These questions will be further explored in
future work.

\subsection*{Acknowledgments}
We thank T.~Adams, H.~Prosper, and M.~Strassler for discussions.
We also thank B.~Tweedie and Z.~Chacko for comments on the manuscript.
The work of CK is supported by DOE grant DE-FG02-96ER50959.
KK is supported by DOE grant DE-FG02-97ER41022 and the Department of Physics
at FSU. TO is supported by a First Year Assistant Professor Award and the
Department of Physics at FSU.

\appendix

\section{Forbidding proton decay}
\Appl{proton}
In \Sec{unification}, we did not require the unification scale to be high enough to
suppress proton decay.  For example, in the benchmark $V+2X$ model, the unification
scale is $M_\text{GUT} \sim 10^{11}\GeV$, significantly lower than the standard GUT
scale $\sim 10^{16}\GeV$. Therefore, it is essential to discuss how proton decay can
be avoided in principle, even though it has no phenomenological relevance at the TeV
scale.

A robust way to avoid the proton decay problem is simply to forbid it by a
symmetry. For example, we can consider ``baryon triality''~\cite{Btriality-1},
$\phi \to \e^{\I\fr{2\pi}{3}(B_\phi -2 Y_\phi)}\phi$, where $B_\phi$ and $Y_\phi$
are the baryon number and hypercharge of a field $\phi$, respectively. Baryon
triality implies that the baryon number can only be violated in units of 3.
Thus, it will absolutely forbid not only proton decay but also neutron-antineutron
oscillation~\cite{Btriality-2}. Alternatively, for forbidden proton decay only,
one could impose ``quark parity'', $q,d^\C,u^\C \to -q,-d^\C,-u^\C$.

Clearly, such symmetries necessarily treat quarks and leptons differently, so
the simplest possibility of embedding quarks and leptons into unified GUT
multiplets (e.g.~$\bar{\mbf{5}} \oplus \mbf{10}$ for $\SU(5)$) does not work.
Instead, quarks and leptons must come from separate multiplets, e.g.:
\beq
  \bar{\mbf{5}}_{d^\C} = \column{d^\C \cr 0}  ,\quad
  \bar{\mbf{5}}_{\ell} = \column{0 \cr \ell}  ,
\eeq
where $\bar{\mbf{5}}_{d^\C}$ and $\bar{\mbf{5}}_{\ell}$ carry the same baryon triality
(or quark parity) quantum numbers as $d^\C$ and $\ell$, respectively. Before discussing
how to promote such incomplete multiplets to fully GUT-invariant multiplets, we would
like to point out that splitting SM matter fields from their heavy ``GUT partners'' is
plausible in the sense that there is already a split multiplet, i.e.~the Higgs doublet,
and whatever mechanism that splits the Higgs from its GUT partner could also work for the
SM matter fields.  Moreover, separating quarks and leptons allows us to trivially incorporate
non-unified mass relations for the 1st and 2nd generations (i.e.~$m_e/m_d, m_\mu/m_s \neq 1$
at the unification scale).

Now, let us discuss how the above ``incomplete'' multiplets can be compatible with a GUT
symmetry. A simple and plausible way to do so is to copy the mechanism~\cite{copyQCD}
of ``incomplete'' multiplets nature already has in the low-energy QCD. QCD undergoes a
symmetry breaking from  $G \equiv \SU(3)_\mrm{L} \otimes \SU(3)_\mrm{R}$ down to
$H\equiv \SU(3)_{\mrm{L} \oplus \mrm{R}}$. Here, the low-mass hadrons form ``incomplete''
multiplets'', i.e.~multiplets of the unbroken subgroup $H$, not those of the full group $G$.
Of course, the theory must be invariant under $G$, which is accomplished by nonlinear
realization~\cite{CCWZ} in terms of the Nambu-Goldstone boson field $\Sg$ transforming as
$\Sg \to g\Sg h^{-1}$ with $g\in G$ and $h\in H$. This permits us to promote any multiplet
of $H$, $\phi_H$, to a full multiplet of $G$, $\phi_H$, as $\phi_G \equiv \Sg \phi_H$.

By analogy, let us consider the following GUT scenario. Imagine a new confining strong
dynamics with a ``flavor symmetry'' $G=\SU(5)$ which undergoes ``chiral symmetry breaking''
$G \to H$ with $H = \SU(3) \otimes \SU(2) \otimes \U(1)$.  $G$ is also weakly gauged,
providing the SM gauge group with a single unified coupling at the $G$ confinement scale
(which therefore can be called the GUT scale).  We then imagine that the split quark and lepton
multiplets as well as the Higgs doublet are ``hadrons'' of the new strong dynamics which form
multiplets of $H$, with appropriate quark parity or baryon triality, etc. They can be promoted
to full $G$ multiplets by nonlinear realization. (If we further integrate in the ``$\rho$ mesons''
into this picture by employing ``hidden local symmetry''~\cite{HLS}, then we essentially obtain
the 2-site moose model considered in \Ref{neal}.)

Constructing an explicit, UV-complete 4D gauge theory that realizes this scenario is beyond
the scope of this paper, but it is straightforward to construct a 5D realization via the
AdS/CFT correspondence~\cite{AdSCFT}, as is done in \Ref{holoGUT}.  The simplest setup would be the Randall-Sundrum
framework~\cite{RS}, i.e., where we take the UV boundary and the bulk to be $G$-symmetric
while the IR boundary to be only $H$-symmetric, by putting a $G$ gauge field in the bulk with
the boundary condition that the $G/H$ gauge bosons vanish at the IR boundary. The split matter
and Higgs multiplets can then be put either in the bulk with the appropriate boundary conditions
to project out the unwanted ``$G$-partner'' states, or simply at the IR boundary as multiplets
of $H$.  With an appropriate symmetry such as quark parity or baryon triality, this setup solves
the proton decay problem, but since the scale associated with the IR boundary is the GUT scale,
there are no observable consequences of the Kaluza-Klein excitations.
(There are, however, RS GUT models with the TeV-scale IR
brane, with split multiplets and a discrete symmetry to forbid proton decay, which do have
experimental consequences at the TeV scale~\cite{RS-GUTs}. The main difference between those
models and ours is that their physics above the TeV scale is a strong conformal dynamics (or
a 5D theory via AdS/CFT) while we assume a perturbative 4D physics up to the GUT scale.)

\section{Effects of higher-dimensional operators}
\Appl{higher-dim-op}
Here we would like to show that adding nonrenormalizable interactions to our Lagrangian
\eq{benchmark} does not alter our results.  In particular, one might worry that a
higher-dimensional operator might lead to a prompt decay of the $X$ even when $\la = 0$,
destroying the R-hadron scenario. Fortunately, this is not the case. The leading
nonrenormalizable interaction that can let $X$ decay is
\beq
  \fr{1}{\La} (X^\C H^*) (q H^*)  \,,
\eeq
where $\La$ is some high scale.  By our assumption, there is no new threshold between the
TeV scale and the unification scale, so $\La$ must be at least $\sim10^{11}\GeV$. To avoid
a suppression from three-body phase space, one of the Higgs fields can be put to its VEV.
Therefore, the $X$ decay length due to this operator is at least
\beq
  \Ga^{-1}
      \sim   \lt[ \fr{m_X v^2}{16\pi (10^{11}\GeV)^2} \rt]^{-1}
      \simeq 3~\text{m} \, \fr{1\TeV}{m_X}   \,.
\eeq
Thus even the most conservative estimate gives a decay length comparable to the dimensions
of the LHC detectors. Therefore, higher-dimensional operators do not upset our conclusions.
Since this $X$ decay is still prompt on a cosmological time scale, setting
$\la = 0$ in the coupling \eq{HdX} is actually allowed cosmologically, as $X$ can decay via the
above operator.


\end{document}